\documentclass[letterpaper, 10 pt, conference]{ieeeconf}  

\IEEEoverridecommandlockouts                              

\usepackage{graphics} 
\usepackage{epsfig} 
\usepackage{mathptmx} 
\usepackage{times} 
\usepackage{amsmath} 
\usepackage{amssymb}  

 \usepackage{tabularx}
 \usepackage{multirow}
 \usepackage{makecell}
 \usepackage{longtable}
 \usepackage{hyperref}
 \usepackage[figuresright]{rotating}

\usepackage[automake]{glossaries}
\usepackage{makeidx}
\makeglossaries
\glsaddall
\newacronym{STPA}{STPA}{Systems Theoretic Process Analysis}
\newacronym{STAMP}{STAMP}{System-Theoretic Accident Model and Processes}
\newacronym{LESs}{LESs}{Learning-Enabled Systems}
\newacronym{ML}{ML}{Machine Learning}
\newacronym{AVs}{AVs}{Autonomous Vehicles}
\newacronym{UUVs}{UUVs}{Unmanned Underwater Vehicles}
\newacronym{UCAs}{UCAs}{Unsafe Control Actions}
\newacronym{UFoI-E}{UFoI-E}{Uncontrolled Flows of Information and Energy}
\newacronym{Bowtie}{Bowtie}{Bow Tie methodology}
\newacronym{FMEA}{FMEA}{Failure Modes and Effects Analysis}
\newacronym{FTA}{FTA}{Fault tree analysis}
\newacronym{CAST}{CAST}{Causal Analysis based on System Theory}
\newacronym{HAZOP}{HAZOP}{Hazard and Operability Analysis}
\newacronym{HILLS}{HILLS}{Hierarchical HAZOP-Like method for LESs}
\newacronym{SOTA}{SOTA}{State of the Art}
\newacronym{CPSs}{CPSs}{Cyber-Physical Systems}
\newacronym{SLIM}{SLIM}{Success Likelihood Index Method}
\newacronym{AHP}{AHP}{Analytic Hierarchy Process}
\newacronym{SPN}{SPN}{Stochastic Petri Nets}
\newacronym{IOU}{IOU}{Intersection Over Union}
\newacronym{AEB}{AEB}{Autonomous Emergency Braking}
\newacronym{DRL}{DRL}{Deep Reinforcement Learning}
\newacronym{NNs}{NNs}{Neuron Networks}

\newcount\Comments  
\usepackage{color}
\definecolor{darkgreen}{rgb}{0,0.5,0}
\definecolor{purple}{rgb}{1,0,1}
\newcommand{\kibitz}[2]{\ifnum\Comments=1\textcolor{#1}{#2}\fi}

\title{\LARGE \bf
STPA for Learning-Enabled Systems: A Survey and A New Practice
}

\author{Yi Qi$^{1}$, Yi Dong$^{1}$, Siddartha Khastgir$^{2}$, Paul Jennings$^{2}$, Xingyu Zhao$^{1,2}$ and Xiaowei Huang$^{1}$
\thanks{*This work has received funding from the European Union’s Horizon 2020 research and innovation programme under grant agreement No 956123.  It is also financially supported by the U.K. EPSRC through End-to-End Conceptual Guarding of Neural Architectures [EP/T026995/1].}
\thanks{$^{1}$University of Liverpool, L69 3BX, U.K. 
        {\tt\small \{yi.qi,yi.dong, xingyu.zhao,xiaowei.huang\}@liverpool.ac.uk}}%
\thanks{$^{2}$University of Warwick, CV4 7AL, U.K.
        {\tt\small \{s.khastgir.1, paul.jennings,xingyu.zhao\}@warwick.ac.uk}}%
}

\usepackage[absolute,overlay]{textpos}

\begin{document}

\begin{textblock*}{20cm}(1cm,1cm)
	\textcolor{red}{Accepted by  the 26th IEEE Int. Conf. on Intelligent Transportation Systems (ITSC'23)}
\end{textblock*}

\maketitle
\thispagestyle{empty}
\pagestyle{empty}

\begin{abstract}


Systems Theoretic Process Analysis (STPA) is a systematic approach for hazard analysis that has been used across many industrial sectors including transportation, energy, and defense. The unstoppable trend of using Machine Learning (ML) in safety-critical systems has led to the pressing need of extending STPA to Learning-Enabled Systems (LESs). Although works have been carried out on various example LESs, without a systematic review, it is unclear how effective and generalisable the extended STPA methods are, and whether further improvements can be made. To this end, we present a systematic survey of 31 papers, summarising them from five perspectives (attributes of concern, objects under study, modifications, derivatives and processes being modelled). Furthermore, we identify room for improvement and accordingly introduce DeepSTPA, which enhances STPA from two aspects that are missing from the state-of-the-practice: (i) Control loop structures are explicitly extended to identify hazards from the data-driven development process spanning the ML lifecycle; (ii) Fine-grained functionalities are modelled at the layer-wise levels of ML models to detect root causes. We demonstrate and compare DeepSTPA and STPA through a case study on an autonomous emergency braking system.

\end{abstract}

\section{INTRODUCTION}


\gls{STPA} is a hazard analysis method based on \gls{STAMP} \cite{leveson2018stpa}, which has been 
used in aerospace \cite{moreira2022stpa}, aviation \cite{salgado2021cybersecurity}, national defence \cite{williams2019system}, nuclear power \cite{rejzek2018use}, as well as other industries \cite{sultana2019hazard}.
In \gls{STAMP}, system safety is regarded as a control problem. Uncontrolled external disturbances, component failures and/or abnormal component interactions can result in system accidents.  System safety is ensured when the control process adheres to safety constraints \cite{leveson2018stpa}.
Based on \gls{STAMP}, \gls{STPA} begins with a control structure model to evaluate each stage of the function within the control loop and then identify the hazards that can impact the system's dynamic behaviour.

Recently, the use of 
\gls{ML} to analyse complex data and integrate them into \gls{CPSs}, known as \gls{LESs}, has become widespread. When applying \gls{LESs} in safety-critical domains, safety assurance is essential to their successful deployment and regulatory compliance, which remains a pressing challenge \cite{BKCF2019,KKB2019,zhao_safety_2020} 
despite recent efforts \cite{10.1007/978-3-030-54549-9_18,10.1145/3453444,hawkins2021guidance,dong_reliability_2022}. Hazard identification is the first and critical step in safety assurance which enables the detection of causes and measures for mitigations of safety risks. Given the popularity of \gls{STPA} for traditional safety-critical systems, there is a growing interest in utilising STPA for LESs. However, there is no dedicated literature review on this emerging research direction, and the following questions are yet to be answered: 

\textit{ Whether the original STPA method requires adaptation to address new characteristics of ML components? How effective is STPA for LESs?  Is there  room for improvement?}

To answer, we first conduct a systematic survey on the use of STPA for LESs to showcase its recent advancements from five perspectives, i.e., attributes of concern, objects under study, modifications to STPA, derivatives of the analysis, and processes being modelled as control loops. While STPA appears to be a promising technique towards safer LESs, the survey suggests two major gaps in the state-of-the-art:

\textbf{Gap 1}: Most studies only consider the system operation process \emph{after} the deployment. For traditional systems without ML components, operational failure information is normally sufficient to identify root causes and mitigations. However, for LESs, the root causes of ML failures can be in the model architecture, training process, or data collection and cannot be fully understood and detected by only observing operational failures. This view of considering all stages in the ML lifecycle has been presented in the academic discussions on the certification process \cite{machinelearningsafetybook,10.1145/3453444} and the policy level discussions on the governance structure \cite{Edwards2022,UKAssurancePolicy}.

\textbf{Gap 2}: In all surveyed studies, STPA works at the software/hardware component level, i.e., the minimal granularity of the functionalities modelled by STPA control structures is an entire component. This is sufficient for traditional hardware/software components whose specifications and implementations are well understood. However, for LESs, locating causes at layer-wise level \textit{inside} ML models (e.g., those arising from inputs to a fully-connected layer and feature maps extracted by a convolutional layer) can be more effective in both identifying and mitigating hazards. This aligns with research efforts to unpack complex ML models through visualisation, explanations, verification and testing.

To bridge these two gaps, we introduce DeepSTPA, a novel extension of STPA for \gls{LESs}. DeepSTPA improves upon STPA in two ways: (i) it explicitly depicts how each stage of the ML lifecycle handles data through control loop structures, to uncover hazards originating from the data-driven development process; (ii) it represents finer functionalities down to the layers of ML models to uncover ``deeper'' root causes for more effective mitigations of hazards. We demonstrate DeepSTPA through a comparative case study on \gls{AEB} systems.

Contributions of this paper include: \textit{i)} A systematic literature survey devoted to STPA and LESs is conducted, which concludes with positive findings and areas for improvement from diverse perspectives.
\textit{ii)} Based on identified problems with the state-of-the-art, a new and adapted way of practicing STPA called DeepSTPA is proposed to explicitly consider ML characteristics. \textit{iii)} We demonstrate DeepSTPA with real-world case studies, including a \gls{DRL} based \gls{AEB} system.


\section{Preliminaries}
\subsection{Learning-Enabled Systems}


\gls{LESs} are systems that utilise \gls{ML} algorithms to continuously improve their performance over time \cite{10.1007/978-3-030-54549-9_18}. These ML models are data-driven and make predictions/decisions, normally being integrated into \gls{CPSs} \cite{sandha2022learning}. 
However, despite their numerous benefits in adapting nonlinearities for complex and high-dimensional tasks, \gls{LESs} are not immune to failures. 

\begin{figure}[htbp]
	\centering
	\includegraphics[width=0.95\linewidth]{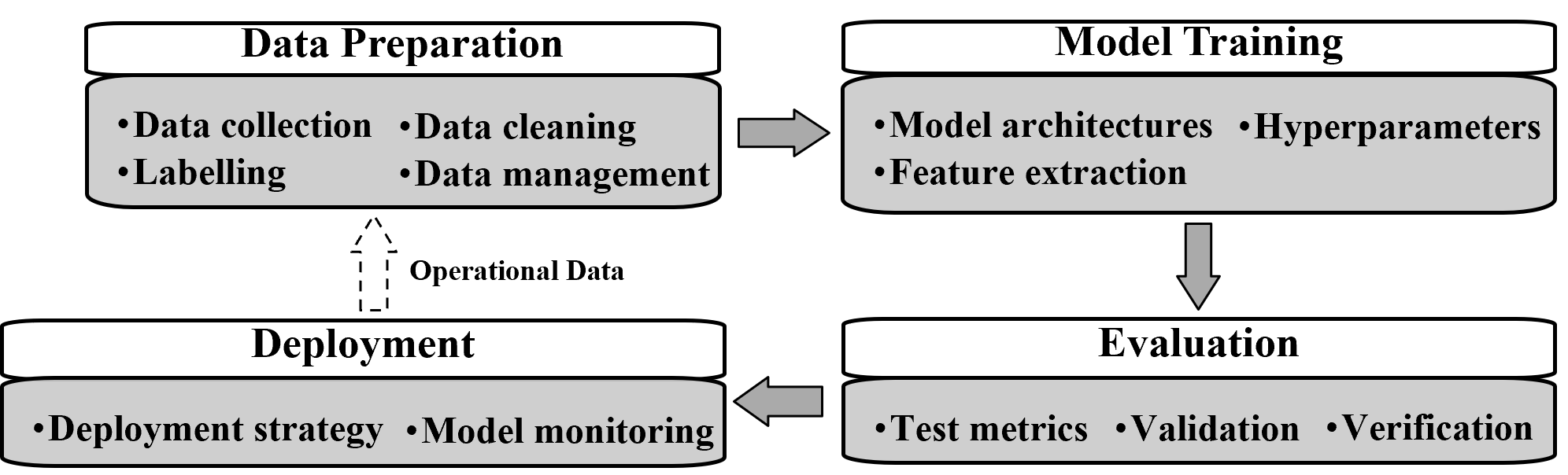}
	\caption{The ML lifecycle comprises four stages---a spiral process model}
	\label{fig_ml_lifecycle}
\end{figure}

A comprehensive ML lifecycle model can be found in \cite{10.1145/3453444}, which identifies assurance desiderata for each stage and reviews existing methods that contribute to achieving the desiderata. An ML lifecycle generally consists of four stages---data preparation, model training, evaluation and deployment, as shown in Fig.~\ref{fig_ml_lifecycle}. Instead of a waterfall process, it is normally a spiral process \cite{10.1145/3453444} that new data gathered during the operation of existing models can be exploited and, where appropriate, new models may be learnt and deployed.

Fig. \ref{fig_ml_lifecycle} illustrates the process of supervised learning, where data preparation involves multiple steps to preprocess the sensor data. For unsupervised learning and reinforcement learning, some of these steps need adaptation. For example, reinforcement learning requires the environment preparation instead of data preparation, which involves creating the simulation scenarios and defining the reward function.

\begin{figure}[htbp]
	\centering
	\includegraphics[width=0.8\linewidth]{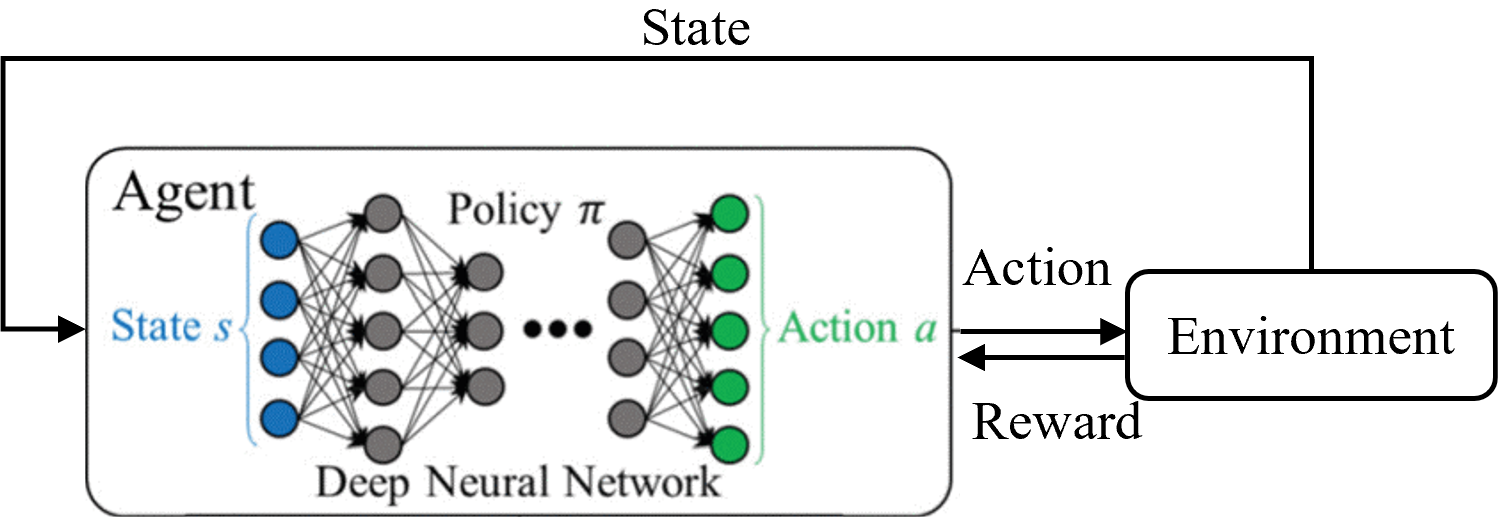}
	\caption{Schematic structure of deep reinforcement learning }
	\label{drlstructure}
\end{figure}

As an example, \gls{DRL} is a subfield of ML that combines deep learning and reinforcement learning \cite{arulkumaran2017deep}, which is adopted in our later case study. Fig.~\ref{drlstructure} displays the architecture of DRL which consists of the environment and agent. In DRL, the agent's objective is to learn a policy that maximises the total reward it accumulates over time. The environment interacts with the agent by providing states, receiving actions from the agent, and offering rewards as feedback. The agent's decisions are guided by the observed states and its policy, and the actions it chooses impact the subsequent states and rewards. Through a combination of exploration and exploitation, the agent learns to make informed decisions to optimise its actions for higher rewards within the environment \cite{arulkumaran2017deep}.
Deep Q Network (DQN) is a specific DRL algorithm that uses \gls{NNs} to approximate Q-values, representing expected future rewards for different actions in a given state. \cite{mnih2013playing}. 
\gls{NNs} usually consist of several layers, each with specific functionalities, e.g., \textbf{convolution layers} implement features extraction, \textbf{pooling layers} preserve the main features while reducing parameters and computation, and \textbf{fully-connect layers} in which each neuron applies a linear transformation to the input vector through a weights matrix.

\subsection{Systems Theoretic Process Analysis}
In \gls{STAMP}, three fundamental concepts from systems theory are used---constraints, hierarchical levels of control, and process models. These concepts provide the foundation for analysing complex systems and understanding the interactions and relationships between various components in a system that can lead to accidents \cite{leveson2011engineering}. 
STPA utilises STAMP's control structures \cite{leveson2011engineering} to find potential risks or hazards of the whole system, and identify causal scenarios and present safety requirements before the system is put into use. Typically, the \gls{STPA} workflow includes the following steps \cite{leveson2018stpa}:
 1) Define Accidents/Hazards. 2) Model Control Structure. 3) Identify Unsafe Control. 4) Identify Causal Scenarios. 5) Derive Safety Requirements.  

The first step is to define some accidents or hazards from a high level. Typically, these are  serious consequences for personnel or equipment that can be anticipated.
The following step is to establish a system control structure that separates the control loop from its components, including the starting point, intermediate processes (e.g., mechanical and software structures), and completion components. Then, potential \gls{UCAs} can be identified using standard \gls{UCAs} form with four categories: \textit{(T1) Not providing the control action leads to a hazard}; \textit{(T2) Providing the control action leads to a hazard}; \textit{(T3) Providing a potentially safe control action but too early, too late, or in the wrong order}; and \textit{(T4) The control action lasts too long or is stopped too soon}. 
The \gls{STPA} then identifies potential risks/hazards by finding \gls{UCAs} and determining the possible causes and scenarios. Finally, the \gls{STPA} provides safety requirements for each potential risk/hazard.





\section{STPA for LESs: A Survey}

\subsection{Scope and Paper Collection}

In the literature search, we adopted Atlas.ti 6.0 \cite{friese2013user}, a qualitative research analysis tool to collect papers. The search function of this survey is presented below, and these keywords only applied to the paper title and abstract.
{\small
\begin{equation}
\label{f1}
\begin{split}
\!\!\! Search & : = \left[\textit{STPA} | \textit{STAMP}\right] + [ (\textit{Learning Enabled Systems}) |\\
 (&\textit{Robotics \& Autonomous Systems})  | (\textit{Machine Learning})]
\end{split}
\end{equation} }\normalsize
where $+$ indicates ``AND'', and $|$ indicates ``OR''.
Moreover, papers, books and thesis were excluded based on the following criteria: i) not published in English; ii) strictly less than four pages; iii) exists duplicated versions; iv) can not be retrieved using IEEE Explore, Google Scholar, Electronic Journal Center, or ACM Digital Library.

We selected 144 papers using search function \eqref{f1}, then filtered out those that only mentioned STPA in the introduction, related work, and future work sections, leaving 59 papers for further analysis. After careful examination, we further narrowed down our selection to 31 papers by removing duplicates and excluding those that did not combine the two themes of STPA and ``learning'' as a single topic despite mentioning both sets of keywords in the main methodology and case studies sections, cf.~Table~\ref{tab_stpa_survey}. It summarises the works from five perspectives---\textit{Attributes of Concern}, \textit{Objects Under Study}, \textit{Modification of STPA}, \textit{Derivatives of the Analysis}, and \textit{Processes being Modeled as Control Loops}.

\begin{table*}
\caption{Surveyed papers (\textbf{NB}, AVs, ASs, Cobots, AMRs, ACs, AUVs, RFSs, MRM, OP and DP represent Autonomous Vehicles, Autonomous Ships, Collaborative Robots, Autonomous Mobile Multi-robots, Automatic Cranes, Autonomous Underwater Vehicles, Robotic Flight Simulator, Multilevel Runtime Monitoring, Operation Process and Development Process, respectively)}
\begin{tabular}[c]{|p{0.7cm}<{\centering}|p{0.7cm}<{\centering}|p{2.3cm}<{\centering}|p{2cm}<{\centering}|p{3.8cm}<{\centering}|p{4cm}<{\centering}|p{1.1cm}<{\centering}|}
\hline
\textbf{Year} & \textbf{List} & \textbf{Attributes of concern} & \textbf{Object under study} & \textbf{Modification of the method} & \textbf{Derivative of the analysis} & \textbf{Process Modeled} \\
\hline
2015 & \cite{abdulkhaleq2015xstampp} & Safety & N/A & Yes (XSTAMPP) & Requirements & OP\\ \hline
2016 & \cite{7527748} & Safety, Privacy & Smart TV & Yes (STPA-priv) & Requirements & OP\\ \hline
2017 & \cite{ABDULKHALEQ201741} & Safety & AVs & No & Requirements, New architectures & OP\\ \hline
2017 & \cite{8054835} & Safety, Privacy & E-health & No & Requirements & OP\\ \hline
2017 & \cite{CHATZIMICHAILIDOU201713} & Safety & Drones & No & Requirements & OP\\ \hline
2018 & \cite{9a073f7edfa248aaa88799b0116c480b} & Safety, Security & AVs & Yes (STPA with six-step model) & Requirements & DP, OP\\ \hline
2018 & \cite{DAKWAT2018130} & Safety & RFSs & Yes (STPA with UPPAAL) & Requirements, New architectures & OP\\ \hline
2019 & \cite{PEREIRA2019302} & Safety, Security & Aeronautics & Yes (STPA-sec) & Requirements & OP\\ \hline
2019 & \cite{sharma_et_al:OASIcs:2019:10338} & Safety, Security & AVs & No & Requirements, Test cases & OP\\ \hline
2019 & \cite{refId0} & Safety & ASs & No & Requirements, Test cases & OP\\ \hline
2020 & \cite{9282673} & Safety, Security & AVs & Yes (STAMP SnS) & Requirements & DP, OP \\ \hline
2020 & \cite{bensaci2020stpa} & Safety & Cobots & Yes (STPA with Bowtie) & Requirements & OP\\ \hline
2020 & \cite{bensaci2020new} & Safety & AMRs & Yes (STPA with FTA) & Requirements & OP\\ \hline
2021 & \cite{9722016} & Safety & ACs & No & Requirements & OP \\ \hline
2021 & \cite{9653486} & Safety & AVs & No & Requirements & OP \\ \hline
2021 & \cite{ADRIAENSEN2021534} & Safety & Cobots & No & Requirements & OP\\ \hline
2021 & \cite{KHASTGIR2021107610} & Safety & AVs & No & Requirements, Test cases & OP\\ \hline
2021 & \cite{guzman2021comparative} & Safety & ASs & Yes (STPA with UFoI-E) & Requirements & OP\\ \hline
2021 & \cite{9582542} & Safety & AVs & Yes (SysML-STPA) & Requirements, New architectures & OP\\ \hline
2021 & \cite{10.1115/1.4051940} & Safety & AVs & No & Requirements & OP\\ \hline
2021 & \cite{DGHAYM2021105139} & Safety & AUVs & Yes (SE-STPA) & Requirements & OP\\ \hline
2021 & \cite{20.500.12210/59063} & Safety & AMRs & Yes (STPA with AHP) & Requirements & OP\\ \hline
2021 & \cite{ZHOU2021108569} & Safety & ASs & Yes (STPA-SynSS) & Requirements & OP\\ \hline
2022 & \cite{Yamada_2022} & Safety & ASs & No & Requirements & OP\\ \hline
2022 & \cite{9985097} & Safety & AVs & No & Requirements & OP\\ \hline
2022 & \cite{10.1007/978-3-031-16245-9_3} & Safety & AMRs, Cobots & No & Requirements, Test cases & OP\\ \hline
2022 & \cite{AHN2022112643} & Safety, Reliability & ASs & Yes (STPA with SLIM) & Requirements & OP\\ \hline
2022 & \cite{10.1007/978-3-031-14835-4_11} & Safety & N/A & Yes (STPA with MRM) & Requirements & OP\\ \hline
2023 & \cite{BENSACI2023109138} & Safety & AMRs & Yes (STPA with SPN) & Requirements & OP\\ \hline
2023 & \cite{10041909} & Safety, Security & AV & No & Requirements, New architectures & OP\\ \hline
2023 & \cite{qi2023safety} & Safety & AV & Yes (STPA with ChatGPT) & Requirements & OP\\ \hline
\end{tabular}
\label{tab_stpa_survey}
\end{table*}




\subsection{Survey Results} 

\paragraph{Attributes of Concern}
Safety is the attribute that \gls{STPA} has traditionally focused on, with the main goal of identifying hazards and enhancing  the safety of the applied system.
With the advancement of \gls{ML}, in addition to safety, security and privacy have emerged as the two attributes of growing interest, which encompass guarding against external system breaches and noticing the privacy implications of system usage. 
This is exemplified by the development of STPA-sec \cite{PEREIRA2019302,9282673,9a073f7edfa248aaa88799b0116c480b,sharma_et_al:OASIcs:2019:10338,10041909} and STPA-priv \cite{8054835,7527748}, indicating the flexibility of \gls{STPA} in terms of accommodating attributes beyond safety. STPA-sec focuses on analysing the vulnerability of the system to external attacks, while STPA-priv highlights the data privacy issues in the system. Reliability is also incorporated in \cite{AHN2022112643} which uses STPA  to analyse human activities that affect the risk of human-machine interaction in probabilistic risk assessment. 

\paragraph{Object Under Study}

Autonomous (ground) vehicles are the most commonly studied object by STPA due to their popularity, complexity and pressing need for safety \cite{9282673,9722016,ABDULKHALEQ201741,sharma_et_al:OASIcs:2019:10338,9a073f7edfa248aaa88799b0116c480b,9653486,KHASTGIR2021107610,9582542,10.1115/1.4051940,DGHAYM2021105139,9985097,10041909,qi2023safety}. Autonomous ships and drones are also studied in \cite{Yamada_2022,refId0,guzman2021comparative,ZHOU2021108569,AHN2022112643} and \cite{CHATZIMICHAILIDOU201713}, respectively. Collaborative robots \cite{bensaci2020stpa,ADRIAENSEN2021534,10.1007/978-3-031-16245-9_3} and Autonomous Mobile Multi-robots \cite{bensaci2020new,20.500.12210/59063,10.1007/978-3-031-16245-9_3,BENSACI2023109138} are also active areas of applying STPA. Moreover, STPA-priv was initially implemented for smart televisions \cite{7527748} before being extended to E-health \cite{8054835}, recognising that privacy is a vital concern for humans. Moreover, STPA-sec has also found use in the Aeronautic industry \cite{PEREIRA2019302}.
STPA appears to be effective in analysing such complex \gls{CPSs}, thanks to its holistic and systematic approach, emphasising causality for early hazard identification.

\paragraph{Modification of the Method}
While STPA can be directly applied on LESs, due to the intricacy of LESs, many studies advocate combining STPA with other safety analysis methods, e.g., the aforementioned STPA-sec and STPA-priv, the combined use of STPA and \gls{CAST} \cite{abdulkhaleq2015xstampp}, a hybrid method for assessing system reliability combines STPA and Success Likelihood Index Method (SLIM) \cite{AHN2022112643}, and the combination of STPA with \gls{UFoI-E} \cite{guzman2021comparative}, \gls{Bowtie} \cite{bensaci2020stpa}, \gls{FTA} \cite{bensaci2020new} and multilevel run-time monitors \cite{10.1007/978-3-031-14835-4_11}. It is also feasible to integrate STPA with hierarchical modelling approaches for intricate systems \cite{9a073f7edfa248aaa88799b0116c480b,9282673}. STPA may jointly analyse with \gls{AHP} \cite{20.500.12210/59063} and \gls{SPN} \cite{BENSACI2023109138} respectively and recommend new safety requirements. The advancements have been made in the integration of STPA with domain-specific safety frameworks for autonomous vehicles and ships, e.g. \cite{9582542,DGHAYM2021105139,ZHOU2021108569}. In \cite{DAKWAT2018130}, STPA is combined with model checking to offer a formal and unambiguous representation of the analysed system. How STPA can make use of Large Language Models like ChatGPT was firstly investigated in \cite{qi2023safety}. 

\paragraph{Derivatives of the Analysis}
Typically, the final step of STPA is to propose safety requirements as a solution to hazards identified. In our survey, it has come to our attention that more diverse use of STPA has been documented.
Specifically, corresponding test cases can be created and different results can be obtained by testing the LESs in varying environments \cite{sharma_et_al:OASIcs:2019:10338,KHASTGIR2021107610,refId0}. Safety issues can be pinpointed at the system architecture design level, thus mitigation strategies based on new architecture can be derived from STPA \cite{9582542,ABDULKHALEQ201741}. 

\paragraph{Processes Modelled as Control Loops}
In most studies, STPA is used to analyse system dynamics during the operation, where interactions between end-users/operators, hardware/software components are modelled as actions and feedback in control loops. However, unlike traditional systems for which operational information may be sufficient for the identification of hazards, causes and mitigations, root causes of ML failures can be located in the data management, ML architecture, training process, etc. Thus, the safety analysis of LESs should also consider activities involving the data owners, providers (who train the ML model), deployers (who deploy the ML model into a larger LES), and the end-users, which requires the modelling of each stage in the ML lifecycle. Moreover, as the ML model becomes more versatile and ``deeper'', fine-grained functionalities at layer-wise level are introduced. To locate the causes from such lower levels and provide effective mitigations for identified hazards, it may be necessary to model how these layers inside the ML model process input data during the operation. To bridge the two gaps we adapted a new way of practicing STPA, DeepSTPA, in the next section.

\section{An Adapted Practice: DeepSTPA}

\begin{figure*}[htbp]
	\centering
	\includegraphics[width=\linewidth]{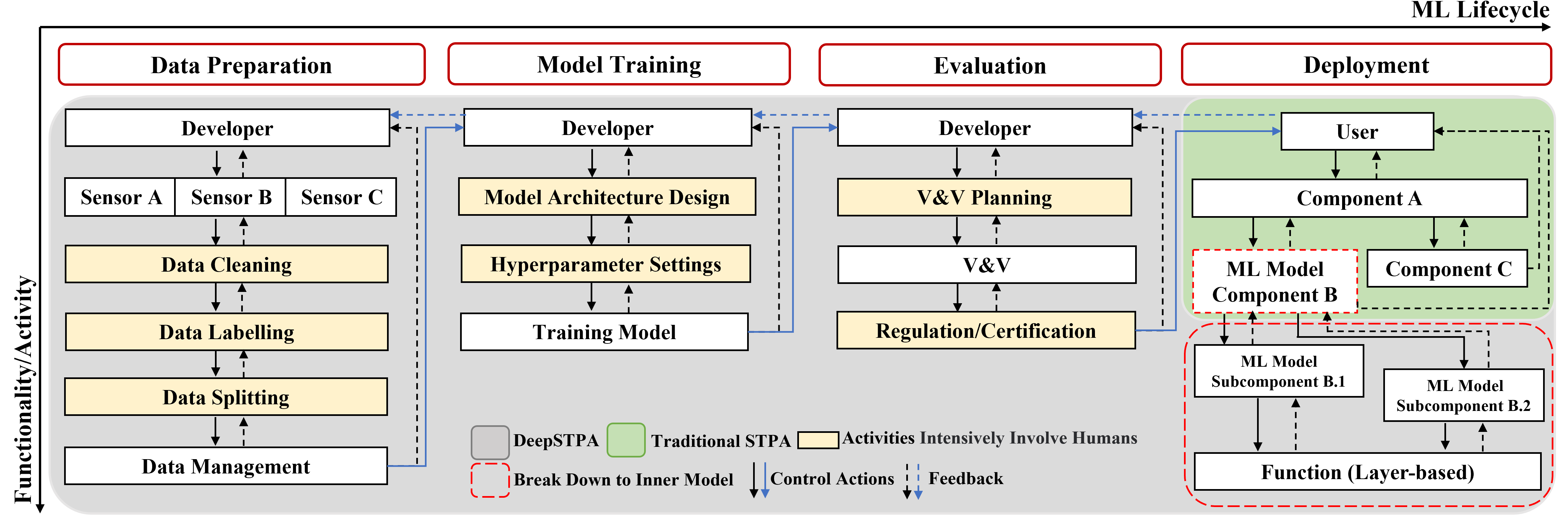}
	\caption{DeepSTPA control loop structures (grey shaded) in addition to the traditional STPA control loop (green shaded)}
	\label{fig_deepstpa_control}
\end{figure*}

We develop an adapted practice called DeepSTPA, 
to cater for LESs with respect to those two gaps we identified in the survey. 
The general framework is depicted in Fig.~\ref{fig_deepstpa_control}---for a given LES, grey shaded areas contain new control loops modelled by DeepSTPA while the green region contains the control loops from the original STPA. 

Specifically, in Fig.~\ref{fig_deepstpa_control}, the horizontal axis signifies the ML lifecycle highlighting how data is being processed, while the vertical axis signifies fine-grained functionalities and development activities. The solid line signifies control action (e.g., human commands, messages and data), and the dashed line represents feedback information. Golden boxes highlight activities involve developers. The red dashed box indicates that the ML component is divided into its inner model, showing finer-grained functionalities at a layer-wise level.

\section{A comparative study with DeepSTPA}\label{sec_deep}
\label{deep_stpa}

DeepSTPA is applied on two case studies---object detection (YOLOv3) on AUVs and a DRL-based AEB system on AVs. Considering the page limit, we only present the latter, with a baseline in \cite{10.1115/1.4051940} for comparative studies, while cf. our project website: \url{https://github.com/YiQi0318/DeepSTPA} for the former.



\subsection{Automatic Emergency Braking Systems}
If a collision is about to occur and the driver takes no action or the action is not fast enough, AEB system will automatically initiate braking. AEB is able to detect potential collisions and activate the braking system to slow down the vehicle to avoid the collision or reduce its impact.
It is a common subsystem on both conventional and autonomous vehicles. A typical AEB system consists of many components, including signal acquisition, calculation, algorithm, and fusion processes, as well as interfaces with electrical and mechanical parts, sensor systems, and more. Recently ML-based AEB solutions are emerging, e.g., in \cite{8317839}.

\subsection{Baseline Results of STPA with AEB systems}

We choose the STPA result of \cite{10.1115/1.4051940} as our baseline, in which an AEB system was analysed. Given the main adaptation of DeepSTPA to STPA is by modelling more comprehensive control loop structures (cf. the grey shaded area compared to the green area in Fig.~\ref{fig_deepstpa_control}), the analysis will primarily concentrate on comparing the modelling control structures---\textit{if more control actions can be modelled, consequently more UCAs, causal scenarios and safety requirements may also be identified}.



The STPA control loop structure yield from \cite{10.1115/1.4051940}, representing a minimum unit AEB system with sensors, controller and actuators, is shown in the green shaded area of Fig.~\ref{fig_deepstpaaeb} (while omitting those example causal factors leading to hazards in the original baseline paper).


\subsection{Results of DeepSTPA applied on DRL-based AEB systems}
\subsubsection{Mishaps Definition}
DeepSTPA's first step resembles STPA, where potential mishaps are listed and accidents are determined by LESs' application scenarios. Since AVs colliding with the assets may cause severe monetary loss or human injured, mishaps (\textbf{M}) considered are:
\begin{itemize}
    \item \textbf{M1}: Damage to the AVs
    \item \textbf{M2}: Damage to the surrounding assets
    \item \textbf{M3}: Human injured
\end{itemize}

\subsubsection{Control Loop Modelling}

\begin{figure*}[htb]
	\centering
	\includegraphics[width=\linewidth]{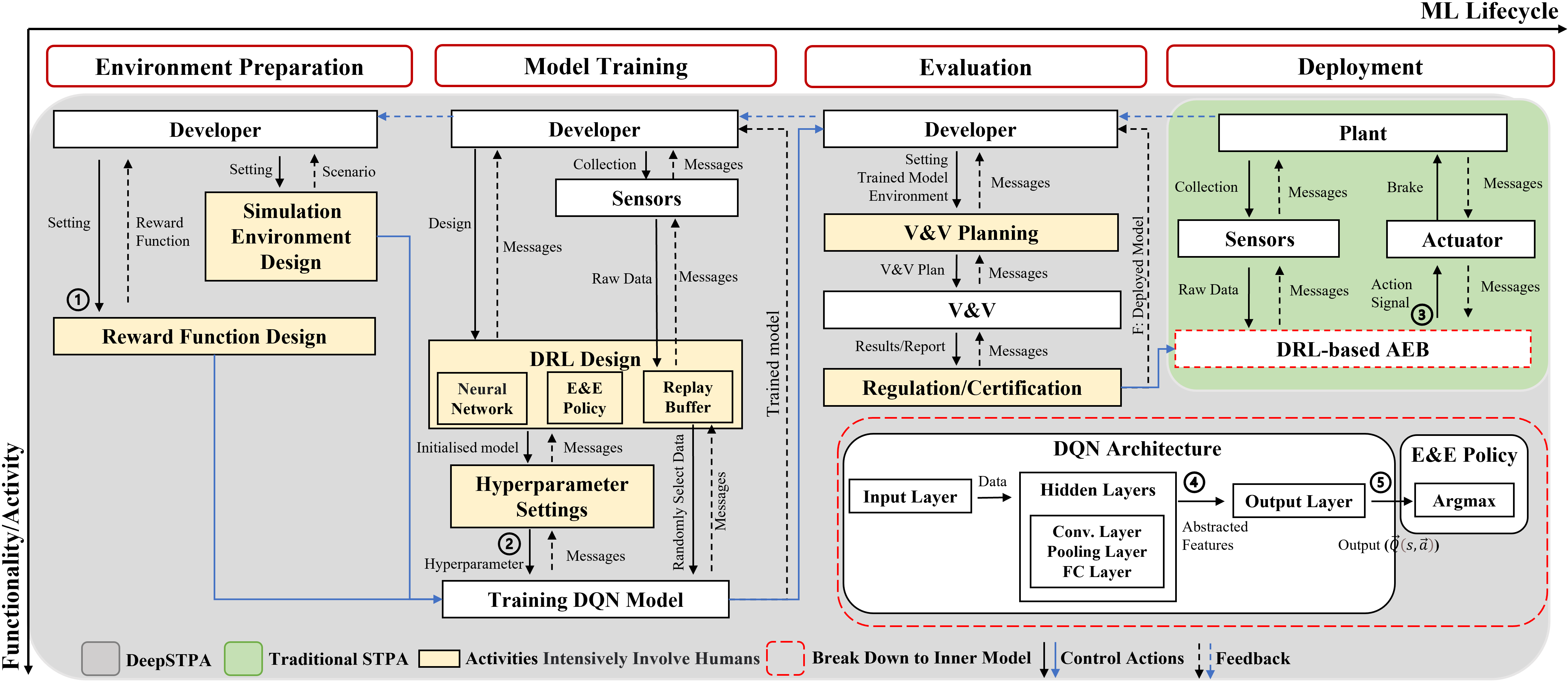}
	\caption{DeepSTPA control loop structures applied on a DRL-based AEB system}
	\label{fig_deepstpaaeb}
\end{figure*}

Fig.~\ref{fig_deepstpaaeb} shows the complete control loop structure modelled by DeepSTPA for the DRL agent of \cite{8317839}, which is a special case of the general framework in Fig.~\ref{fig_deepstpa_control}. 
Specfically, the first stage in the ML lifecycle (i.e., data preparation) is replaced with ``environment preparation''. This adaptation depends mainly on the type of models used in the ML component. The developer will design the reward function and simulation scenarios to prepare for training the model in the next stage (solid blue line across stages).
During the model training stage, it is important to consider 
the architecture design of the NN, the trade-off between exploration and exploitation (E\&E) policy, and configuration of the replay buffer. 
After obtaining the trained model and setting up the verification and validation (V\&V) plan, the V\&V steps are carried out in either simulation or real environments, at the evaluation stage. Then, the ML model can be deployed into a LES. The traditional STPA method can be used to analyse potential risks and hazards (green area) at this operation stage. However, DeepSTPA also explores deeper into the functionality/activity dimension. That is, given the core model of the DRL-based AEB is the DQN model, we further develop control loop structures (red dashed box) to represent layer-wise functionalities. Inputs are passed through the hidden layers for feature extraction, parameter reduction, and computation. The output layer selects the appropriate action signal to send to the actuator based on the E\&E policy.


\subsubsection{Unsafe Control Actions}
There are 4 parts consisting of a hazardous control action of each UCA, \textit{Source Controller}, \textit{Type} (with the 4 aforementioned categories \textit{T1}-\textit{T4}),  \textit{Control Action} and \textit{Context}. 
According to \cite{leveson2018stpa}, the four categories in \textit{Type} are \textit{complete} in terms of defining UCAs. To be more pertinent, DeepSTPA introduces subcategories to indicate new deviations representing ML characteristics applied under the Type \textit{T2}, includeing
``Wrong'', ``Invalid'', ``Incomplete'' and ``Perturbed'',
e.g., ``Wrong setting and NN design'' is a human error of ML model developers. 

In the dimension of ML lifecycle, developers may make human errors in development activities, e.g. (UCA index corresponds to numbers labeled on control actions in Fig.~\ref{fig_deepstpaaeb}):
\begin{itemize}
    \item \textbf{UCA-1} Incomplete reward function design
    \item \textbf{UCA-2} Wrong hyperparameter setting
\end{itemize}

During the environment preparation stage, the developer is responsible for designing a reward function that effectively communicates the mission goals to the agent (including penalties for collisions or undesired actions). However, human errors can occur, such as neglecting secondary goals that the agent should fulfill. This incomplete reward function design can result in poor performance of the trained model, leading to incorrect action signal outputs from the DRL-based AEB component. These failures can have serious consequences (\textbf{M1}, \textbf{M2}, and \textbf{M3}). UCA-2 is another example highlighting a human error in the model training stage that may affect system-level safety eventually. Note, without modelling control structures from the ML lifecycle dimension, the original STPA cannot identify such UCAs.

If potential risks or hazards are found in a certain component during the deployment stage, DeepSTPA can support detailed safety analysis, tracing back (via blue arrows in Fig.~\ref{fig_deepstpaaeb}) to development stages in the ML lifecycle for that component (\textbf{UCA-1} and \textbf{UCA-2}). One example is the passing of the reward function and simulation scenarios from the ``environment preparation'' stage to the ``model training'' stage, and then placing the trained model in the system after the V\&V evaluation stage. This means the ML lifecycle stages formed a control loop structure, allowing us to loop back from the operation process to the ML component development process.

One common UCA related to DRL-based AEB component is \textbf{UCA-3}: DRL-based AEB component does not provide the correct action signal (\textit{T1}). \textbf{UCA-3} can be easily identified using STPA, as it is apparent from the operational control loop structure (green area in Fig.~\ref{fig_deepstpaaeb}). DeepSTPA is an extension of the original STPA that allows for a more detailed analysis that can be further broken down into the internal layers of the ML model. For instance, in the case of a DRL-based AEB system, the core model is the DQN model, and DeepSTPA can help identify that the DQN model does not provide an action signal:
\begin{itemize}
    \item \textbf{UCA-4} The kernel size or stride is too large in the convolutional layer, resulting in a limited amount of information being extracted by the convolutional layer in real-world scenarios\footnote{Kernel size refers to the window size over the input, and a larger size extracts less information causing performance degradation. Stride determines how many pixels the window should move, and a higher stride can lead to more limited feature extraction.}. 
    \item \textbf{UCA-5} There is the gap between the estimated Q value through $argmax$ and the actual maximum Q value\footnote{The $argmax$ function selects the maximum value of ${\textbf{Q}}(s, \textbf{a})$ from the NN and outputs its corresponding action signal to the next component.}.
\end{itemize}

\textbf{UCA-3} 
can be further broken down to the inner ML model (red dashed box in Fig.~\ref{fig_deepstpaaeb}), which may be due to the DQN model not providing the correct action signal. The analysis revealed that the potential risks and hazards at the layer-wise level, as reflected in \textbf{UCA-4} and \textbf{UCA-5}, can cause the model to output incorrect action signals, ultimately affecting the entire component's output and system-level safety.

\subsubsection{Causal Scenarios}

In general, STPA considers two causal scenarios (\textbf{C}): the reasons behind UCAs and the factors contributing to improper/unexecuted control actions. Referring to the aforementioned \textbf{UCA-5}, we present some example causal scenarios (while a more complete list would require DRL experts' knowledge, as expected in STPA \cite{DAKWAT2018130}):
\begin{itemize}
    \item \textbf{C1}: The model's generalisation ability is inadequate\footnote{Generalisation enables good performance on new data beyond training.}.
    \item \textbf{C2}: There exists the explore-exploit dilemma\footnote{It arises explore-exploit dilemma when there is uncertainty about the rewards associated with different actions or strategies. }.
    \item \textbf{C3}: Sample efficiency is not good\footnote{When the sample efficiency is not good, it means that the model requires a large number of training samples to achieve satisfactory performance.}.
    \item \textbf{C4}: It may involve a sparse reward situation\footnote{Sparse rewards refer to instances where the agent receives limited feedback or reward signals from the environment during training.}.
\end{itemize}

\begin{table*}[hbt!]
\caption{Comparison attribute list (adapted from \cite{10.1115/1.4051940}) and results}
\centering
\label{tab3}
\scalebox{0.95}{
\begin{tabular}{|l|l|l|l|}
\hline
Attributes                    & Descriptions & STPA       & DeepSTPA\\ 
\hline \hline
Identify hazards              & Comprehensiveness of identified hazards &  Partially &Comprehensively \\ \hline
Identify causes               & Comprehensiveness of causes of identified hazards   &  Partially & Comprehensively        \\ \hline
Hazard causal factors         & Type of identified hazard causal factors&  Hardware, Software & Hardware, Software, ML Model\\ \hline \hline
Life-cycle phase              & Method can be applied in which phase of the product life-cycle &  Operation phase & Development phase, Operation phase\\ \hline
Time and cost                 & A time and cost required for safety analysis with the method       & Low  &High       \\ \hline
Complexity/difficulty         & Relative complexity/difficulty of the method& Low  &High\\
\hline
\end{tabular}
}
\end{table*}

\subsubsection{Safety Requirements}

From the defined mishaps, UCAs, and causal scenarios, finally a set of safety requirements (\textbf{R}) can be derived to enhance the overall safety. One of the common safety requirements involves the creation of safety project checklists or the implementation of consistency checks for specific components. Additionally, standards may be formulated to ensure adherence to safety protocols and guidelines. To address \textbf{UCA-5}, effective solutions include:
\begin{itemize}
    \item \textbf{R1}: Apply data augmentation techniques during the model training process\footnote{Data augmentation diversifies and increases training data to improve the model's ability to generalize and handle real-world variations.}.
    \item \textbf{R2}: Maintain a balance between exploration and exploitation in reinforcement learning, it can be employed state-of-the-art methods like the $\epsilon$-greedy approach\footnote{The $\epsilon$-greedy approach is a strategy that enables the agent to explore different actions while also exploiting the currently known optimal actions.}.
    \item \textbf{R3}: Incorporate an experience replay\footnote{Replaying past experiences during training helps the model learn from diverse sets of data, enhancing efficiency and stability.}.
    \item \textbf{R4}: Lead into the reward shaping to relieve the sparse reward situation\footnote{ Reward shaping is a technique that involves adding additional reward signals to guide the RL agent during training. These additional rewards are designed to provide more informative feedback to the agent.}.
\end{itemize}
These safety requirements help mitigate risks and ensure that safety considerations are effectively incorporated into the operation process.

In summary, DeepSTPA follows the same methodology as STPA---it retains the five basic steps, but enhances the 2nd step significantly by considering layer-wise functionalities inside ML models and development activities spanning the ML lifecycle (cf. the grey shaded area in Fig.~\ref{fig_deepstpa_control}). Thanks to  such more comprehensive modelling of control loop structures, DeepSTPA can provide causal scenarios and safety requirements for ML specific UCAs that complements STPA. 

\subsection{Comparison Methodology and Results}
We reuse the comparison methodology from \cite{10.1115/1.4051940} (designed for a diverse range of safety analysis methods), but reduce the set of attributes to suit our specific goal on comparing DeepSTPA to STPA. Specifically, we assess and compare DeepSTPA and STPA according to the attributes list in Table~\ref{tab3}. The six attributes collectively encompasses two aspects: \textit{analysis results} comparison and \textit{analysis process} comparison. In Table \ref{tab3}, the first three rows correspond to the analysis results comparison, while the remaining rows pertain to the analysis process comparison.


The summarised resutls of the comparative study between STPA and DeepSTPA are presented in Table \ref{tab3}. The table reveals that only a part of the hazards can be identified in STPA, while DeepSTPA provides a more comprehensive analysis results (hazards, causes and mitigations). This is non-surprising, given that DeepSTPA can model more control actions (during both the development process and fine-grained ML functionalities in the operation) and consequently more UCAs, causal scenarios and safety requirements can be derived.
However, the utilisation of DeepSTPA comes at the cost of increased complexity and longer time requirements compared to STPA.

\section{Related Work}

STPA was introduced as a replacement for traditional safety analysis methods, e.g., HAZOP \cite{lawley1974operability}, in the context of handling complex systems \cite{e2019system}. Researchers explored the combination of STPA with other techniques and its adaptation to a broad range of domains. Rodriguez and Diaz \cite{rodriguez2014new} study the integration of STPA with functional models for process hazard analysis, while Kondo et al. \cite{kondo2018application} adapt the STAMP framework by modifying its terminology to incorporate risk analysis for industrial control systems and cybersecurity. 
Pasman's work \cite{pasman2015risk} offers a systematic approach to risk control and explores new and improved processes, as well as various risks and resilience analysis tools. 
There have been surveys conducted on STPA. Harkleroad et al. \cite{harkleroad2013review} review the general properties of \gls{STPA}, summarising that \gls{STPA} views risk assessment as a top-down control issue instead of some commonly used bottom-up component reliability approaches.
Patriarca et al. \cite{PATRIARCA2022105566} have recently summarised the history and current status of STAMP, which encompasses \gls{STPA} as one of its related techniques, highlighting the application areas of surveyed papers. To the best of our knowledge, none of the existing surveys have specifically focused on LESs like ours. 
Driven by the same motivation as DeepSTPA (i.e., a new way of doing safety analysis for LESs), in \cite{qi_hierarchical_2022}, authors develop a new method by featuring HAZOP a hierarchical structure, while we are extending STPA which is arguably more effective for complex systems like LESs.


\section{CONCLUSIONS}


In this paper, we conduct a survey of 31 papers selected from recent literature of applying STPA to LESs. Based on the survey results, gaps in the state-of-the-art are identified for which we propose a new practice, DeepSTPA, adapting STPA from the dimensions of ML development process and fine-grained functionalities of layers inside ML models. As demonstrated by the comparative case study on AVs, we conclude that DeepSTPA is more effective than STPA when applied to LESs, which can be reflected through the more comprehensive safety analysis results.


 \addtolength{\textheight}{-2mm}   









\bibliographystyle{IEEEtran}
\bibliography{mybibliography}

\begin{thebibliography}{10}
\providecommand{\url}[1]{#1}
\csname url@rmstyle\endcsname
\providecommand{\newblock}{\relax}
\providecommand{\bibinfo}[2]{#2}
\providecommand\BIBentrySTDinterwordspacing{\spaceskip=0pt\relax}
\providecommand\BIBentryALTinterwordstretchfactor{4}
\providecommand\BIBentryALTinterwordspacing{\spaceskip=\fontdimen2\font plus
\BIBentryALTinterwordstretchfactor\fontdimen3\font minus
  \fontdimen4\font\relax}
\providecommand\BIBforeignlanguage[2]{{%
\expandafter\ifx\csname l@#1\endcsname\relax
\typeout{** WARNING: IEEEtran.bst: No hyphenation pattern has been}%
\typeout{** loaded for the language `#1'. Using the pattern for}%
\typeout{** the default language instead.}%
\else
\language=\csname l@#1\endcsname
\fi
#2}}

\bibitem{leveson2018stpa}
N.~G. Leveson and J.~P. Thomas, ``{STPA} handbook,'' \emph{USA}, 2018.

\bibitem{moreira2022stpa}
G.~Moreira, D.~R. Pleffken, C.~Cerqueira, and W.~Santos, ``{STPA} analysis over
  the earlier phases of brazilian aerospace products life cycle using {OPM},''
  in \emph{ICMAE}, 2022, pp. 465--471.

\bibitem{salgado2021cybersecurity}
M.~L. Salgado and M.~S. de~Sousa, ``Cybersecurity in aviation: the {STPA}-sec
  method applied to the {TCAS} security,'' in \emph{LADC}, 2021.

\bibitem{williams2019system}
A.~D. Williams, ``System theoretic process analysis ({STPA}): Overview of
  sandia uses to address national security problems.'' 2019.

\bibitem{rejzek2018use}
M.~Rejzek and C.~Hilbes, ``Use of {STPA} as a diverse analysis method for
  optimization and design verification of digital instrumentation and control
  systems in nuclear power plants,'' \emph{NED}, pp. 125--135, 2018.

\bibitem{sultana2019hazard}
S.~Sultana, P.~Okoh, S.~Haugen, and J.~E. Vinnem, ``Hazard analysis:
  Application of stpa to ship-to-ship transfer of lng,'' \emph{JLPPI}, 2019.

\bibitem{BKCF2019}
R.~Bloomfield, H.~Khlaaf, P.~R. Conmy, and G.~Fletcher, ``Disruptive
  innovations and disruptive assurance: Assuring machine learning and
  autonomy,'' \emph{Computer}, 2019.

\bibitem{KKB2019}
P.~Koopman, A.~Kane, and J.~Black, ``Credible autonomy safety argumentation,''
  in \emph{27th {Safety}-{Critical} {Systems} {Symp.}}, 2019.

\bibitem{zhao_safety_2020}
X.~Zhao, A.~Banks, J.~Sharp, V.~Robu, D.~Flynn, M.~Fisher, and X.~Huang, ``A
  {Safety} {Framework} for {Critical} {Systems} {Utilising} {Deep} {Neural}
  {Networks},'' in \emph{SafeComp'20}, ser. {LNCS}, vol. 12234, 2020.

\bibitem{10.1007/978-3-030-54549-9_18}
E.~Asaadi, E.~Denney, and G.~Pai, ``Quantifying assurance in learning-enabled
  systems,'' in \emph{SafeComp'20}, ser. {LNCS}, vol. 12234, 2020.

\bibitem{10.1145/3453444}
R.~Ashmore, R.~Calinescu, and C.~Paterson, ``Assuring the machine learning
  lifecycle: Desiderata, methods, and challenges,'' \emph{ACM Comput. Surv.},
  vol.~54, no.~5, 2021.

\bibitem{hawkins2021guidance}
R.~Hawkins, C.~Paterson, C.~Picardi, Y.~Jia, R.~Calinescu, and I.~Habli,
  ``Guidance on the assurance of machine learning in autonomous systems
  ({AMLAS}),'' \emph{arXiv preprint arXiv:2102.01564}, 2021.

\bibitem{dong_reliability_2022}
Y.~Dong, W.~Huang, V.~Bharti, V.~Cox, A.~Banks, S.~Wang, X.~Zhao, S.~Schewe,
  and X.~Huang, ``Reliability {Assessment} and {Safety} {Arguments} for
  {Machine} {Learning} {Components} in {System} {Assurance},'' \emph{ACM Trans.
  Embed. Comput. Syst.}, vol.~22, no.~3, 2023.

\bibitem{machinelearningsafetybook}
X.~Huang, G.~Jin, and W.~Ruan, \emph{Machine Learning Safety}.\hskip 1em plus
  0.5em minus 0.4em\relax Springer, 2023.

\bibitem{Edwards2022}
L.~Edwards, ``Regulating ai in europe: four problems and four solutions,''
  \emph{Retrieved March}, vol.~15, 2022.

\bibitem{UKAssurancePolicy}
{Department for Digital, Culture, Media and Sport, U.K.}, ``Establishing a
  pro-innovation approach to regulating {AI},'' Tech. Rep., 2022.

\bibitem{sandha2022learning}
S.~S. Sandha, \emph{Learning-enabled Cyber-Physical Systems: Challenges and
  Strategies}.\hskip 1em plus 0.5em minus 0.4em\relax University of California,
  Los Angeles, 2022.

\bibitem{arulkumaran2017deep}
K.~Arulkumaran, M.~P. Deisenroth, M.~Brundage, and A.~A. Bharath, ``Deep
  reinforcement learning: A brief survey,'' \emph{IEEE Signal Processing
  Magazine}, vol.~34, no.~6, pp. 26--38, 2017.

\bibitem{mnih2013playing}
V.~Mnih, K.~Kavukcuoglu, D.~Silver, A.~Graves, I.~Antonoglou, D.~Wierstra, and
  M.~Riedmiller, ``Playing atari with deep reinforcement learning,''
  \emph{arXiv preprint arXiv:1312.5602}, 2013.

\bibitem{leveson2011engineering}
N.~Leveson, \emph{Engineering a Safer World: Systems Thinking Applied to
  Safety}, ser. Engineering systems.\hskip 1em plus 0.5em minus 0.4em\relax MIT
  Press, 2011.

\bibitem{friese2013user}
S.~Friese, ``User’s manual for {ATLAS}. ti 6.0, {ATLAS}. ti scientific
  software development,'' \emph{Berlin: GmbH}, 2013.

\bibitem{abdulkhaleq2015xstampp}
A.~Abdulkhaleq and S.~Wagner, ``{XSTAMPP}: an extensible {STAMP} platform as
  tool support for safety engineering,'' \emph{STAMP WS}, 2015.

\bibitem{7527748}
S.~S. Shapiro, ``Privacy risk analysis based on system control structures:
  Adapting system-theoretic process analysis for privacy engineering,'' in
  \emph{IEEE Security and Privacy Workshops (SPW)}.\hskip 1em plus 0.5em minus
  0.4em\relax IEEE, 2016.

\bibitem{ABDULKHALEQ201741}
A.~Abdulkhaleq, D.~Lammering, S.~Wagner, J.~Röder, N.~Balbierer, L.~Ramsauer,
  T.~Raste, and H.~Boehmert, ``A systematic approach based on {STPA} for
  developing a dependable architecture for fully automated driving vehicles,''
  \emph{Procedia Eng}, vol. 179, pp. 41--51, 2017.

\bibitem{8054835}
K.~Mindermann, F.~Riedel, A.~Abdulkhaleq, C.~Stach, and S.~Wagner,
  ``Exploratory study of the privacy extension for system theoretic process
  analysis (stpa-priv) to elicit privacy risks in ehealth,'' in \emph{IEEE 25th
  Int. Req. Engi. Conf. workshops (REW)}.\hskip 1em plus 0.5em minus
  0.4em\relax IEEE, 2017, pp. 90--96.

\bibitem{CHATZIMICHAILIDOU201713}
M.~M. Chatzimichailidou, N.~Karanikas, and A.~Plioutsias, ``Application of
  {STPA} on small drone operations: A benchmarking approach,'' \emph{Procedia
  Engineering}, 2017, 4th European STAMP Workshop.

\bibitem{9a073f7edfa248aaa88799b0116c480b}
G.~Sabaliauskaite, L.~Liew, and J.~Cui, ``Integrating autonomous vehicle safety
  and security analysis using {STPA} method and the six-step model,''
  \emph{Int. J. on Adv. in Security}, vol.~11, pp. 160--169, 2018.

\bibitem{DAKWAT2018130}
A.~L. Dakwat and E.~Villani, ``System safety assessment based on {STPA} and
  model checking,'' \emph{Safety Science}, vol. 109, 2018.

\bibitem{PEREIRA2019302}
D.~P. Pereira, C.~Hirata, and S.~Nadjm-Tehrani, ``A {STAMP}-based ontology
  approach to support safety and security analyses,'' \emph{Journal of
  Information Security and Applications}, vol.~47, pp. 302--319, 2019.

\bibitem{sharma_et_al:OASIcs:2019:10338}
S.~Sharma, A.~Flores, C.~Hobbs, J.~Stafford, and S.~Fischmeister, ``{Safety and
  Security Analysis of AEB for L4 Autonomous Vehicle Using STPA},'' in
  \emph{Workshop on ASD}, vol.~68, 2019, pp. 5:1--5:13.

\bibitem{refId0}
{Rokseth, B\o{}rge}, {Haugen, Odd Ivar}, and {Utne, Ingrid Bouwer}, ``Safety
  verification for autonomous ships,'' \emph{MATEC Web Conf.}, 2019.

\bibitem{9282673}
T.~Kaneko, N.~Yoshioka, and R.~Sasaki, ``{STAMP S\&S}: Safety \& security
  scenario for specification and standard in the society of {AI/IoT},'' in
  \emph{QRS-C}, 2020, pp. 168--175.

\bibitem{bensaci2020stpa}
C.~Bensaci, Y.~Zennir, D.~Pomorski, F.~Innal, Y.~Liu, and C.~Tolba, ``{STPA}
  and {Bowtie} risk analysis study for centralized and hierarchical control
  architectures comparison,'' \emph{Alexandria Engineering Journal}, vol.~59,
  no.~5, pp. 3799--3816, 2020.

\bibitem{bensaci2020new}
C.~Bensaci, Y.~Zennir, and D.~Pomorski, ``A new approach to system safety of
  human-multi-robot mobile system control with {STPA} and {FTA},''
  \emph{Algerian Journal of Signals and Systems}, pp. 79--85, 2020.

\bibitem{9722016}
W.~Zhang, X.~Meng, J.~Wang, T.~Li, Q.~Shan, and F.~Teng, ``Safety analysis of
  automatic crane trolley running system based on {STAMP}/{STPA},'' in
  \emph{ICCSS}, 2021.

\bibitem{9653486}
B.~Li, S.~Shang, and Y.~Fu, ``The application of stpa in the development of
  autonomous vehicle functional safety,'' in \emph{ICAA}, 2021, pp. 863--868.

\bibitem{ADRIAENSEN2021534}
A.~Adriaensen, L.~Pintelon, F.~Costantino, G.~D. Gravio, and R.~Patriarca, ``An
  {STPA} safety analysis case study of a collaborative robot application,''
  \emph{IFAC}, vol.~54, no.~1, pp. 534--539, 2021.

\bibitem{KHASTGIR2021107610}
S.~Khastgir, S.~Brewerton, J.~Thomas, and P.~Jennings, ``Systems approach to
  creating test scenarios for automated driving systems,'' \emph{Reliability
  Engineering \& System Safety}, vol. 215, p. 107610, 2021.

\bibitem{guzman2021comparative}
N.~H.~C. Guzman, J.~Zhang, J.~Xie, and J.~A. Glomsrud, ``A comparative study of
  {STPA}-extension and the {UFoI-E} method for safety and security
  co-analysis,'' \emph{Reliability Engineering \& System Safety}, 2021.

\bibitem{9582542}
A.~Ahlbrecht and O.~Bertram, ``Evaluating system architecture safety in early
  phases of development with {MBSE} and {STPA},'' in \emph{ISSE}, 2021.

\bibitem{10.1115/1.4051940}
L.~Sun, Y.-F. Li, and E.~Zio, ``{Comparison of the HAZOP, FMEA, FRAM, and STPA
  Methods for the Hazard Analysis of Automatic Emergency Brake Systems},''
  \emph{ASCE-ASME J Risk and Uncert in Engrg Sys Part B Mech Engrg}, vol.~8,
  no.~3, 2021.

\bibitem{DGHAYM2021105139}
D.~Dghaym, T.~S. Hoang, S.~R. Turnock, M.~Butler, J.~Downes, and B.~Pritchard,
  ``An {STPA}-based formal composition framework for trustworthy autonomous
  maritime systems,'' \emph{Safety Science}, 2021.

\bibitem{20.500.12210/59063}
C.~Bensaci, Y.~Zennir, D.~Pomorski, F.~Innal, and Y.~Liu, ``Distributed vs.
  hybrid control architecture using {STPA} and {AHP} - application to an
  autonomous mobile multi-robot system,'' \emph{Int. Journal of Safety and
  Security Engin.}, vol.~11, pp. 1--12, 3 2021.

\bibitem{ZHOU2021108569}
X.~Zhou, Z.~Liu, F.~Wang, and Z.~Wu, ``A system-theoretic approach to safety
  and security co-analysis of autonomous ships,'' \emph{Ocean Engineering},
  vol. 222, p. 108569, 2021.

\bibitem{Yamada_2022}
T.~Yamada, M.~Sato, R.~Kuranobu, R.~Watanabe, H.~Itoh, M.~Shiokari, and
  T.~Yuzui, ``Evaluation of effectiveness of the {STAMP} / {STPA} in risk
  analysis of autonomous ship systems,'' \emph{JPCS}, 2022.

\bibitem{9985097}
G.~Ge, L.~Sun, and Y.~F. Li, ``A systematic approach to develop an autopilot
  sensor monitoring system for autonomous delivery vehicles based on the {STPA}
  method,'' in \emph{ISSREW}, 2022, pp. 318--325.

\bibitem{10.1007/978-3-031-16245-9_3}
L.~Buysse, D.~Vanoost, J.~Vankeirsbilck, J.~Boydens, and D.~Pissoort, ``Case
  study analysis of {STPA} as basis for dynamic safety assurance
  of autonomous systems,'' in \emph{EDCC}, 2022.

\bibitem{AHN2022112643}
S.~I. Ahn, R.~E. Kurt, and O.~Turan, ``The hybrid method combined {STPA} and
  {SLIM} to assess the reliability of the human interaction system to the
  emergency shutdown system of lng ship-to-ship bunkering,'' \emph{Ocean
  Engineering}, vol. 265, p. 112643, 2022.

\bibitem{10.1007/978-3-031-14835-4_11}
S.~Gautham, G.~Bakirtzis, A.~Will, A.~V. Jayakumar, and C.~R. Elks,
  ``{STPA}-driven multilevel runtime monitoring for {In}-time hazard
  detection,'' in \emph{SafeComp'20}, 2022, pp. 158--172.

\bibitem{BENSACI2023109138}
C.~Bensaci, Y.~Zennir, D.~Pomorski, F.~Innal, and M.~A. Lundteigen, ``Collision
  hazard modeling and analysis in a multi-mobile robots system transportation
  task with {STPA} and {SPN},'' \emph{Reliability Engineering \& System
  Safety}, vol. 234, p. 109138, 2023.

\bibitem{10041909}
S.~Ghosh, A.~Zaboli, J.~Hong, and J.~Kwon, ``An integrated approach of threat
  analysis for autonomous vehicles perception system,'' \emph{IEEE Access},
  vol.~11, pp. 14\,752--14\,777, 2023.

\bibitem{qi2023safety}
Y.~Qi, X.~Zhao, and X.~Huang, ``Safety analysis in the era of large language
  models: A case study of {STPA} using {ChatGPT},'' \emph{arXiv 2304.01246},
  2023.

\bibitem{8317839}
H.~Chae, C.~M. Kang, B.~Kim, J.~Kim, C.~C. Chung, and J.~W. Choi, ``Autonomous
  braking system via deep reinforcement learning,'' in \emph{IEEE 20th Int.
  Conf. on Intelligent Transportation Systems}, 2017.

\bibitem{lawley1974operability}
H.~Lawley, ``Operability studies and hazard analysis,'' \emph{Chem. Eng.
  Prog.}, vol.~70, no.~4, 1974.

\bibitem{e2019system}
S.~R. e~Silva, ``System theoretic process analysis: A literature survey on the
  approaches used for improving the safety in complex systems,'' in
  \emph{Information Systems for Industry 4.0}, 2019, pp. 97--114.

\bibitem{rodriguez2014new}
M.~Rodr{\'\i}guez and I.~D{\'\i}az, ``A new functional systems theory based
  methodology for process hazards analysis,'' \emph{CACE}, pp. 703--708, 2014.

\bibitem{kondo2018application}
S.~Kondo, H.~Sakashita, S.~Sato, T.~Hamaguchi, and Y.~Hashimoto, ``An
  application of {STAMP} to safety and cyber security for {ICS},'' in
  \emph{Computer Aided Chemical Engineering}, 2018, vol.~44, pp. 2335--2340.

\bibitem{pasman2015risk}
H.~J. Pasman, \emph{Risk analysis and control for industrial processes-gas, oil
  and chemicals: a system perspective for assessing and avoiding
  low-probability, high-consequence events}, 2015.

\bibitem{harkleroad2013review}
E.~Harkleroad, A.~Vela, and J.~Kuchar, ``Review of systems-theoretic process
  analysis ({STPA}) method and results to support {NextGen} concept assessment
  and validation,'' \emph{Proj. Report: ATC-427 MIT}, 2013.

\bibitem{PATRIARCA2022105566}
R.~Patriarca, M.~Chatzimichailidou, N.~Karanikas, and G.~{Di Gravio}, ``The
  past and present of system-theoretic accident model and processes ({STAMP})
  and its associated techniques: A scoping review,'' \emph{Safety Science},
  vol. 146, p. 105566, 2022.

\bibitem{qi_hierarchical_2022}
Y.~Qi, P.~M. Ryan, W.~Huang, X.~Zhao, and X.~Huang, ``A {Hierarchical}
  {HAZOP}-{Like} {Safety} {Analysis} for {Learning}-{Enabled} {Systems},'' in
  \emph{{AI} {Safety} at {IJCAI}-22}, vol. 3215.\hskip 1em plus 0.5em minus
  0.4em\relax Vienna, Austria: CEUR, 2022, p.~10.

\end{thebibliography}

\end{document}